\documentclass[reprint,showpacs,preprintnumbers, API,prb]{revtex4-1}
\usepackage{graphicx}
\usepackage{dcolumn}
\usepackage{bm}

\preprint{API/123-QED}
\begin{document}
\title{Substrate-limited helical edge states}
\author{B. S. Kandemir}
\email{kandemir@science.ankara.edu.tr}
\author{S. Atag}
\address{Department of Physics, 
Ankara University, Faculty of Sciences, 06100, Tando\u{g}an-Ankara,Turkey}
\date{\today}

\begin{abstract}
We derived analytical results for the gapless edge states of two-dimensional topological insulators in the presence of electron-surface optical (SO) phonon interaction due to substrates.  We followed an analytical algorithm, called  Lee-Low-Pines variational approximation in the conventional polaron theory, to examine the substrate induced effects on both bulk and edge states of a two dimensional topological insulator within the frame work of Bernevig-Hughes-Zhang (BHZ) model. By implementing this algorithm, we propose a novel phonon-dressed BHZ Hamiltonian which allows one  to investigate the effects of various substrates not only on bulk states but also on the associated   gapless helical edge states (HESs). We found that both the bulk and  HESs are significantly renormalized in the momentum space due to the substrate-related polaronic effects. The model we developed here clarifies which subtrates favor the  HESs of quantum spin Hall  system and  which are not.   Correspondingly, our work demonstrates that the substrate related polaronic effects have significant role on the emergence of HESs. In other words, we show that SO phonons due to substrates modify the electronic band topology of topological insulators together with the associated HESs and therefore they can be used to tune quantum phase transitions between topological insulators and non-topological ones.
\end{abstract}

\pacs{73.43.-f, 72.25.Dc,85.75.-d}

\maketitle

\section{Introduction}
Following the first model for quantum Hall effect in the absence of an external magnetic field suggested by Haldene\cite{Haldane1988},
the quantum spin Hall (QSH) phase was proposed as a new state of matter by Kane and Mele\cite{KaneandMele2005}
for graphene system. This QSH system shows an energy gap in the bulk, while it has gapless helical edge states (HESs) with different spins moving in opposite directions.
These gapless HESs are topologically protected by time-reversal symmetry, and they are robust to any perturbations. Their first realistic
theoretical model were predicted by Bernevig \textit{et al}\cite{Bernevig2006},
and soon after they were observed in semiconductor $\mathrm{H} \mathrm{g} \mathrm{T} \mathrm{e}/\mathrm{C} \mathrm{d} \mathrm{T} \mathrm{e}$ quantum wells (QWs) by K\"{o}nig \textit{et al}\cite{Konig2007}.
Later, similar effect arising in Type-II semiconductor QWs made from $\mathrm{I}\mathrm{n}\mathrm{A}\mathrm{s}/\mathrm{G}\mathrm{a}\mathrm{S}\mathrm{b}/\mathrm{A}\mathrm{l}\mathrm{S}\mathrm{b}$ was predicted by Liu \textit{et al}\cite{Liu2008}. Following these pioneering works, there has been a significant interest in studying the exotic properties of QSH effect \cite{Wu2006,Sheng2006,Xu2006,Fu2006,
Onada2007,Fukui2007,Obuse2007,Murakami2007,Qiao2008,Dai2008,FuandKane2008,Qi2009,Yu2010}. However, up till now, apart from the experimental realizations of this effect in these QW systems, its achivement on an appropriate substrate has not been experimentally realized.

It is expected that, when a QSH system is situated on a polar substrate, interaction of the carries of the QSH system with the field induced by surface modes of the dielectric substrate
leads to inevitable effects. In particular, the formation of the HESs of the QSH system is affected by these interactions taken place at the interface
of the substrate and the QSH system. Such a kind of interaction strongly modifies the single particle properties of the system under consideration, leading
to many-body renormalization of the relevant parameters. In fact, the interaction of electrons with the surface optical (SO) phonons of the substrate is
a well-established many body problem since the works of Sak\cite{Sak1972}, Wang
and Mahan\cite{Wang&Mahan1972,Mahan1974}. It is also well-known that, for instance, in graphene,
it is responsible for the modification of many physical properties such as the renormalization of Fermi velocity\cite{Hwang&Sarma2013}, enhanced intra- and inter-band magnetooptical absorption peaks\cite{Scharf2013}.
Thus, to understand their effects on a QSH system, we develop here an analytical method within the frame work of Lee-Low-Pines (LLP)\cite{LLP1953} approximation
in the polaron  theory to propose a novel phonon- dressed BHZ model which comprises  the substrate induced effects on both bulk and edge states. 

Although the QSH phase depends on the universal topological characteristics of the sytem, its emergence in a topological material depends crucially on material spesific
parameters, particularly, on the symmetries of the substrate system upon which topological materials are grown. Indeed, very recently, it is demonstrated
that, to control the relevant orbitals in a two-dimensional (2D) QSH insulators, and thus to create large-gap QSH systems in monolayer-substrate composites, substrates play
decisive roles in the engineering of such materials\cite{Reis2017}. As a matter of fact, it is theoretically shown  that, in  room temperature, bismuthene on $\mathrm{S}\mathrm{i}\mathrm{C}$ substrate is one of the most probable candidates for QSH materials. Our
model developed here not only clarifies why $\mathrm{Si}\mathrm{C}$ substrate favors the edge states of QSH system, but also makes some predictions on which substrates are most
suitable for the QSH system and which are not. Correspondingly, we show that SO phonons due to substrates modify the electronic band topology of topological insulators together with the associated HESs and therefore they can be used to tune quantum phase transitions between topological insulators and non-topological ones. Our claims are also compatible with the predictions of Garate\cite{Garate2013}. He shows that deformation coupling to longitudinal acoustic phonons can alter the topological properties of Dirac insulators.
To date, there have been  already numerous theoretical studies to deal with the effects of deformation potential coupling to longitudinal acoustic phonons on band topology of 3D topological insulators\cite{Giraud2011,Li2012,Parente2013}, topological insulator thin films\cite{Giraud2012} and  HgTe/CdTe quantum wells\cite{Saha2014}(including coupling to nonpolar optical phonons) .

The paper is organized as follows. In Section II and Section III, we present our main results for both bulk and edge state dispersions, respectively, and discuss them in detail. Section IV  ends with a brief conclusion.

\section{Phonon-Dressed BHZ Model}
In the presence of electron-SO phonon interaction, the effective four-band Hamiltonian which was proposed by Bernevig \textit{et
al} \cite{Bernevig2006} in order to QSH effect for  $\mathrm{H} \mathrm{g} \mathrm{T} \mathrm{e}/\mathrm{C} \mathrm{d} \mathrm{T} \mathrm{e}$ QWs can be written as 
\begin{equation}\mathcal{H}_{\mathrm{2D}} (\mathbf{k}) =\mathcal{H}_{\mathrm{B H Z}} (\mathbf{k}) +\mathcal{H} (\mathbf{k}) \mathbf{I}_{4}. \label{1}
\end{equation}
Here, $\mathcal{H}_{\mathrm{B H Z}}$ is $4 \times 4$ Hamiltonian for QSH effect, and is given by 
\begin{equation}\mathcal{H}_{\mathrm{B H Z}} (\mathbf{k}) =\left (\begin{array}{cc}\text{H}(\mathbf{k}) & 0 \\
0 & \text{H}^{ \ast } ( -\mathbf{k})\end{array}\right ), \label{2}
\end{equation}
where $\text{H}(\mathbf{k}) =\epsilon _{k} \mathbf{I}_{2} +d^{a} (k) \sigma _{a}$  is a $2 \times 2$ Hamiltonian  with $\mathbf{I}_{2}$ and $\sigma _{a}$  being $2 \times 2$ unit matrix and  Pauli matrices, respectively. For small k's, $\epsilon _{k} =C -D (k_{x}^{2} +k_{y}^{2})$, $d^{1} =A k_{x}$, $d^{2} =A k_{y}$, and $d^{3} =M -B (k_{x}^{2} +k_{y}^{2})$ together with the material parameters $A$, $B$, $C$, $D$ and $M$, that all depend on the QW geometry. For the QW thickness $d =7.0 \mathrm{n}\text{} \mathrm{m}$ , they  are given as \ $A =3.645\; \mathrm{e} \mathrm{V} \mathring{A}$ ($\hbar  \upsilon _{F}$), $B = -68.6\; \mathrm{e} \mathrm{V}\mathring{A}^{2}$, $D = -51.2\; \mathrm{e} \mathrm{V}\mathring{A}^{2}$, and $M = -0.010\; \mathrm{e} \mathrm{V}$\cite{Konig2008}. It should be noted that,  the upper-left block of Eq.~(\ref{1}), i.e., H$(\mathbf{k})$, which is for spin up, and is related to the lower-right  one  which is for spin down, by
time-reversal symmetry, so it is convenient to focus on the H$(\mathbf{k})$   for the rest of the paper.

The Hamiltonian in Eq.~(\ref{1}),  $\mathcal{H} (\mathbf{k})$, is the sum of Hamiltonians of the free SO-phonons and their coupling to the electron, respectively, and it is taken into account diagonal in the helicity of the Dirac electrons due to the high symmetry of the $\Gamma$ point\cite{Parente2013}. It can be written as 

\begin{equation}\mathcal{H} (\mathbf{k}) =\sum _{\mathbf{q}}\hbar  \omega  b_{\mathbf{q}}^{ \dag } b_{\mathbf{q}} +\sum _{\mathbf{q}}\left [\mathrm{M}_{\mathbf{q}} (z) e^{i \mathbf{q} \cdot \mathbf{r}} b_{\mathbf{q}} +\text{H.C.}\right ] \label{3}
\end{equation}where $\mathbf{r}$ is the 2D position vector of the electron in $xy$-plane , $b_{\mathbf{q}}^{ \dag } (b_{\mathbf{q}})$ is the \ creation (annihilation) operators for a SO phonon of frequency $\omega $ and wave vector $\mathbf{q}$.  $\mathrm{M}_{\mathbf{q}} (z)$ is the interaction amplitude of electrons with SO phonons of the substrate, and its spatial dependence is given
by\cite{Mahan1974} 
\begin{eqnarray*}
\mathrm{M}_{\mathbf{q}} (z) =i \sqrt{g} \frac{e^{ -q z}}{\sqrt{q}} 
\end{eqnarray*}
where $g$\ is the coupling parameter defined by $g =$\ $2 \pi  \hbar  \omega $\ $e^{2} \beta /S$ and $z$ is the distance of the electron from the surface of the substrate. Here, $S$ is the area of the surface, $e$ is the free electron charge together with $\beta  =(\epsilon _{0} -\epsilon _{\infty })/(\epsilon _{0} +1) (\epsilon _{\infty } +1)$ where $\epsilon _{0}$ and $\epsilon _{\infty }$ are low- and high-frequency dielectric constants of the substrate subsystem. Our Fr{\"o}hlich type Hamiltonian for 2D topological insulators given by Eqs.~(\ref{1}-\ref{3}) describes the electrons trapped at the interface between topological material and  the  substrate due to SO phonons of the substrates. The last term in  Eq.~(\ref{3}) contains phonon creation (annihilation) operators $b_{\mathbf{q}}^{ \dag } (b_{\mathbf{q}})$ linearly, and thus it needs to be diagonalized. 

This can be realized by two successive transformations within the framework of LLP\cite{LLP1953} theory. The first unitary transformation
\begin{eqnarray*}
\mathrm{U}_{1} =\exp  \left ( -i \mathbf{r}\cdot\sum _{\mathbf{q}}\hbar  \mathbf{q} b_{\mathbf{q}}^{ \dag } b_{\mathbf{q}}\right ) 
\end{eqnarray*}
eliminates the electron coordinates $\mathbf{r}$ from the interaction Hamiltonian.  Applying  the transformation $\mathrm{U}_{1}$ on  $b_{\mathbf{q}}$ and $\mathbf{p}$   yields $\mathrm{U}_{1}^{ -1} b_{\mathbf{q}} \mathrm{U}_{1} =b_{\mathbf{q}} \exp  ( -i \mathbf{q} \cdot \mathbf{r})$ and $\widetilde{\mathbf{p}} =\mathrm{U}_{1}^{ -1} \mathbf{p} \mathrm{U}_{1} =\mathbf{p} -\hbar  \sum _{\mathbf{q}}\mathbf{q} b_{\mathbf{q}}^{ \dag } b_{\mathbf{q} }$, respectively, so we can write the transformed Hamiltonian $\overline{\text{H}} (\mathbf{k})=\mathrm{U}_{1}^{ -1} \text{H}(\mathbf{k}) \mathrm{U}_{1}$ as 
\begin{widetext}
\begin{eqnarray}
\overline{\text{H}} (\mathbf{k})&=&\left [M -B \left (\mathbf{k} -\sum _{\mathbf{q}}\mathbf{q} b_{\mathbf{q}}^{ \dag } b_{\mathbf{q}}\right )^{2}\right ] \sigma _{z}  +A \mathbf{\sigma } \cdot \left (\mathbf{k} -\sum _{\mathbf{q}}\mathbf{q} b_{\mathbf{q}}^{ \dag } b_{\mathbf{q}}\right )  \nonumber\\
&+&\left\lbrace \left [C -D \left (\mathbf{k} -\sum _{\mathbf{q}}\mathbf{q} b_{\mathbf{q}}^{ \dag } b_{\mathbf{q}}\right )^{2}\right ]+\sum _{\mathbf{q}}\hbar  \omega  b_{\mathbf{q}}^{ \dag } b_{\mathbf{q}} +\sum _{\mathbf{q}}[\mathrm{M}_{\mathbf{q}} (z) b_{\mathbf{q}} +\text{H.C.}]\right\rbrace 
\mathbf{I}_{2} .\label{6}
\end{eqnarray}
\end{widetext}
where   $\mathbf{\sigma } =(\sigma_{x}, \sigma_{y} )$.  Since, the electron-SO phonon interaction part of the Hamiltonian given by Eq.~(\ref{6})
is still non-diagonal in phonon coordinates, we impose the second LLP transformation, to generate coherent boson states from the phonon vacuum $\vert 0 >_{\text{PH}}$, given by
\begin{eqnarray*}
\mathrm{U}_{2} =\mathrm{e} \mathrm{x} \mathrm{p} \left [\sum _{\mathbf{q}}\left (f_{\mathbf{q}} b_{\mathbf{q}}^{ \dag } -f_{\mathbf{q}}^{ \ast } b_{\mathbf{q}}\right )\right ] 
\end{eqnarray*}
which shifts the phonon coordinates by an amount of $f_{q}$, i.e., $\mathrm{U}_{2}^{ -1} b_{\mathbf{q}} \mathrm{U}_{2} =b_{\mathbf{q}} +f_{\mathbf{q}}$. Here, $f_{q}$($f_{q}^{\ast}$) is the variational function to be determined. In terms of the transformed operators, Eq.~(\ref{6})
can be written as $\widetilde{\text{H}} = U_{2}^{ -1} \overline{\text{H}} (\mathbf{k})U_{2} = \text{H}^{(0)} + \text{H}^{(1)} + \text{H}^{(2)}$ .
While $\text{H}^{(1)}$ and $\text{H}^{(2)}$ contains terms with single creation and annihilation terms as well as bilinear ones such as $b_{\mathbf{q}}^{ \dag } b_{\mathbf{q}}$ which all disappear when they are applied to vacuum $\vert 0 >_{\text{PH}}$. The explicit forms of $\text{H}^{(1)}$ and $\text{H}^{(2)}$ are given in Appendix A. $\text{H}^{(0)}$ consists of only the  terms free from phonon operators  whose diagonal matrix components are given as
\begin{widetext}
\begin{eqnarray}
	\text{H}_{11}^{(0)} &  = & C+ M -(B +D) \left [\mathbf{k}^{2} +\sum _{\mathbf{q}}\mathbf{q}^{2} \vert f_{\mathbf{q}}\vert ^{2} -2 \mathbf{k} \cdot \sum _{\mathbf{q}} \mathbf{q} \vert f_{\mathbf{q}}\vert ^{2} +\left (\sum _{\mathbf{q}}\mathbf{q} \vert f_{\mathbf{q}}\vert ^{2}\right )^{2}\right ] \nonumber\\
 &  &  +\sum _{\mathbf{q}}\left [\hbar  \omega  \vert f_{\mathbf{q}}\vert ^{2} +\mathrm{M}_{\mathbf{q}} (z) f_{\mathbf{q}} +\mathrm{M}_{\mathbf{q}}^{ \ast } (z) f_{\mathbf{q}}^{ \ast }\right ],\label{5x}\\
 \text{H}_{22}^{(0)} &  = &  C -M +(B -D) \left [\mathbf{k}^{2} +\sum _{\mathbf{q}}\mathbf{q}^{2} \vert f_{\mathbf{q}}\vert ^{2} -2 \mathbf{k} \cdot \sum _{\mathbf{q}} \mathbf{q} \vert f_{\mathbf{q}}\vert ^{2} +\left (\sum _{\mathbf{q}}\mathbf{q} \vert f_{\mathbf{q}}\vert ^{2}\right )^{2}\right ] \nonumber  \\
 &  &  +\sum _{\mathbf{q}}\left [\hbar  \omega  \vert f_{\mathbf{q}}\vert ^{2} +\mathrm{M}_{\mathbf{q}} (z) f_{\mathbf{q}} +\mathrm{M}_{\mathbf{q}}^{ \ast } (z) f_{\mathbf{q}}^{ \ast }\right ],\label{5xx}
\end{eqnarray}
\end{widetext}
together with non-diagonal ones
\begin{eqnarray}
\text{H}_{12}^{(0)} &=& A \left (k_{ +} -\sum _{\mathbf{q}}q_{ +} \vert f_{\mathbf{q}}\vert ^{2}\right )\nonumber\\
\text{H}_{21}^{(0)} &=& A \left (k_{ -} -\sum _{\mathbf{q}}q_{ -} \vert f_{\mathbf{q}}\vert ^{2}\right ),\label{5xxx}
\end{eqnarray}
where $k_{ \pm } =k_{x} \pm i k_{y}$ and $q_{ \pm } =q_{x} \pm i q_{y}$. Since taking the expectation value of the transformed Hamiltonian $\widetilde{\text{H}}$ by the phonon vacuum state $\vert 0 >_{\text{PH}}$ yields   $ <0\vert \widetilde{\text{H}}\vert 0 >_{\text{PH}}=\text{H}^{(0)}$,  the variation of $\text{H}^{(0)}$ with respect $f_{\mathbf{q}}$ and $f_{\mathbf{q}}^{ \ast }$  leads to
\begin{equation}f_{\mathbf{q}} = -\frac{\mathrm{M}_{\mathbf{q}}^{ \ast } (z)}{\hbar  \omega  \pm B_{ \pm } \left [ -2 \mathbf{k} \cdot \mathbf{q} +\mathbf{q}^{2} +2 \mathbf{q} \cdot \left (\sum _{\mathbf{q}^{^{ \prime }}}\mathbf{q}^{^{ \prime }} \vert f_{\mathbf{q}^{^{ \prime }}}\vert ^{2}\right )\right ]} \label{8}
\end{equation}
and its complex conjugate, respectively.  In fact, this functional minimization procedure of $\text{H}^{(0)}$ corresponds exactly to eliminate  the large part of the residual Hamiltonian given by Eq.~(\ref{A1}), i.e., the part  that includes the phonon operators linearly. This can   be easily verified  that $\text{H}^{(1)}$ vanishes if  $f_{\mathbf{q}}$( $f_{\mathbf{q}}^{ \ast }$ ) satisfies Eq.~(\ref{8}).  Thus, the rest of the Hamiltonian can now be solved exactly if and only if a formal solution to Eq.~(\ref{8}) can be found by solving the implicit functional relations among $f_{\mathbf{q}}$( $f_{\mathbf{q}}^{ \ast }$ ). This can easily be done by following the conventional procedure from the LLP theory. The only preferred direction in the system is the direction of
momentum vector, i.e., $\mathbf{k}$, thus, due to the symmetry rules, so  $\sum _{q^{ \prime }}\mathbf{q}^{ \prime } \vert f_{\mathbf{q}^{ \prime }}\vert ^{2}$ should be differ from $\mathbf{k}$ by a scalar, 
\begin{equation}
	\sum _{\mathbf{q}^{ \prime }}\mathbf{q}^{ \prime } \vert f_{\mathbf{q}^{ \prime }}\vert ^{2} =\eta  \mathbf{k} \label{11}
\end{equation}
that can be solved selfconsistently to minimize the energy of the system by following the common steps in LLP theory. Therefore, it can be easily verified that a solution of the selfconsistent equation for $\eta$ in Eq.~(\ref{11})  may be written as 
\begin{equation}
f_{\mathbf{q}} = -\frac{\mathrm{M}_{\mathbf{q}}^{ \ast } (z)}{\hbar  \omega  +\left \vert B_{ +}\right \vert [\mathbf{q}^{2} -2 \mathbf{k} \cdot \mathbf{q}(1 -\eta )]}\label{12}
\end{equation}
which allows to minimize the total  energy of the system. Substituting Eq.~(\ref{12}) into Eq.~(\ref{11}),
and replacing the summation over $\mathbf{q}$  by $\mathbf{q}$ integral yields 
\begin{eqnarray}
	\eta\mathbf{k}=\frac{1}{4 \pi }\hbar\omega e^{2}\beta\int d^{2}\mathbf{q}\frac{e^{ -2 q z}}{q} 
   \frac{\mathbf{q}}{[\hbar  \omega  +\vert B_{ +}\vert [\mathbf{q}^{2} -2 \mathbf{k} \cdot \mathbf{q}(1 -\eta ]]^{2}}\text{.}\nonumber\\
   \label{16} 
\end{eqnarray}
The integral over $\mathbf{q}$ in Eq.~(\ref{16}) can be analytically evaluated  for slow electrons, $k \ll q_{p} =\sqrt{\hbar  \omega /\vert B_{ +}\vert }$. It should be noted that our small $k$ approximation is compatible with the BHZ model  which describes well only  the
states for small  $k$'s, particularly for the valence band \cite{Krishtopenko2018}.
After, by multiplying both sides of Eq.~(\ref{16}) with $\mathbf{k}$, we first expand the integrand as power series of $\mathbf{k}$, and then keep only the terms up to order $k^{2}$,  it is straightforward to show that the resultant equation 
\begin{eqnarray*}
	\eta  =\frac{1}{4 \sqrt{\pi }} \alpha _{0} \left (1 -\eta \right ) G_{1 ,3}^{3 ,1} \left (\overline{z}^{2}\left \vert \begin{array}{rrr}\, &  -\frac{1}{2} & \, \\
0 & \frac{1}{2} & \frac{3}{2}\end{array}\right .\right )
\end{eqnarray*}
solves $\eta $ as $\eta  =\alpha /\left (1 +\alpha \right )$. This is formaly equivalent to the one obtained from conventinal LLP theory for the bulk polaron, but  with different $\alpha$ composition
\begin{equation}\alpha  =\frac{1}{4 \sqrt{\pi }} \alpha _{0} G_{1 ,3}^{3 ,1} \left (\overline{z}^{2}\left \vert \begin{array}{rrr}\, &  -\frac{1}{2} & \, \\
0 & \frac{1}{2} & \frac{3}{2}\end{array}\right .\right )\label{15}
\end{equation}
with $\alpha _{0} =e^{2} \beta /\sqrt{\hbar  \omega  \vert B_{ +}\vert }$, and $G_{1 ,3}^{3 ,1}$ is the Meijer G-function. The  $\alpha$  in Eq.~(\ref{15}) can be regarded as a position dependent electron-SO phonon coupling parameter in analogy to the bulk polaron theory. Consequently, the diagonal and non-diagonal matrix elements of $\text{H}^{(0)}$ defined by  Eqs.~(\ref{5x}-\ref{5xxx}) can be rewritten as
\begin{widetext}
\begin{eqnarray}
\text{H}^{(0)}_{1 1} &  = &C+ M -B \left (1 -\eta \right )^{2} \mathbf{k}^{2} -C^{01} -(D-D^{01}) \left (1 -\eta \right )^{2} \mathbf{k}^{2} \nonumber  \\
\text{H}^{(0)}_{2 2} &  = & C-M +B \left (1 -\eta \right )^{2} \mathbf{k}^{2} -C^{02} -(D+D^{02}) \left (1 -\eta \right )^{2} \mathbf{k}^{2},\label{19x}
\end{eqnarray}
\end{widetext}
\begin{eqnarray}
\text{H}^{(0)}_{1 2} &  = & A \left (1 -\eta \right ) k_{ +} \nonumber  \\
\text{H}^{(0)}_{2 1} &  = & A \left (1 -\eta \right ) k_{ -},\label{19xx} 
\end{eqnarray}
respectively,
where $C^{01}$, $C^{02}$, $D^{01}$ and $D^{02}$ are all  functions of the parameters of the substrate material as well as material parameters of the topological insulator, and their explicit expressions are given in Appendix B. Thus, by rearranging
the matrix elements of $\text{H}^{(0)}$  in Eqs.~(\ref{19x}-\ref{19xx}-) , we arrive at our new phonon-dressed BHZ Hamiltonian for the  upper-left block as
\begin{widetext}
\begin{equation}
\text{H}^{(0)} \left (\mathbf{k}\right ) =\left [\begin{array}{cc}C^{1} -D^{1}\, \mathbf{k}^{2} +M- B^{1}\, \mathbf{k}^{2} & A^{1}\, k_{ +} \\
A^{1}\, k_{ -} & C^{2} -D^{2}\, \mathbf{k}^{2} -M+ B^{1}\, \mathbf{k}^{2}\end{array}\right ]\label{18a}
\end{equation}
\end{widetext}
with the new phonon-dressed material parameters  $A^{1} =A \left (1 -\eta \right )$, $C^{i} =C -C^{0 i}$, $D^{i} = (D \mp D^{0 i}) \left (1 -\eta \right )^{2}$ (where plus sign is for $i =2$, and minus sign for $1$, respectively), and $B^{1} =B \left (1 -\eta \right )^{2}$.  Subsequently, the bulk energy spectrum of our new phonon-dressed BHZ model, i.e., $E=E(\mathbf{k})$ , can then be found by solving the eigenvalue equation for the upper-left block $\text{H}^{(0)}\Psi _{\uparrow } \left (\mathbf{k}\right )=E(\mathbf{k})\Psi _{\uparrow } \left (\mathbf{k}\right )$ as
\begin{equation}
E_{ \pm } =\mathcal{C} -\mathcal{D}\, \mathbf{k}^{2} \pm [\mathcal{M}^{2} +\left (\mathcal{A}^{2} -2 \mathcal{M}\, \mathcal{B}\right ) \mathbf{k}^{2} +\mathcal{B}^{2} \mathbf{k}^{4}]^{1/2} \label{18}
\end{equation}
where our new phonon-dressed material parameters are given by
\begin{eqnarray}\mathcal{A} &  = & A\, \left (1 -\eta \right ), \nonumber \\
\mathcal{B} &  = & \left [B -\frac{1}{2} \left (D^{01} +D^{02}\right )\right ] \left (1 -\eta \right )^{2}, \nonumber \\
\mathcal{C} &  = & C -\frac{1}{2} \left (C^{01} +C^{02}\right ), \nonumber \\
\mathcal{D} &  = & \left [D +\frac{1}{2} \left (D^{02} -D^{01}\right )\right ] \left (1 -\eta \right )^{2},\nonumber  \\
\mathcal{M} &  = & M +\frac{1}{2} \left (C^{02} -C^{01}\right ).\label{19}
\end{eqnarray}
Eq.~(\ref{18})  is
the the key result of this section, and includes phonon-dressed material parameters given by Eq.~(\ref{19}),
They are all the functions of substrate parameters $\beta $ and $\hbar  \omega $ as well as $z$ through Eqs.~(\ref{B1}-\ref{B5})
in Appendix B, including material parameters of the topological insulator. 
Therefore, both bulk and edge state solutions of Eq.~(\ref{18a}) can obtained in the  standard way but with modified or phonon-dressed material parameters defined by Eq.~(\ref{19}).

\section{Helical Edge States}

In this section, the edge states from the phonon-dressed BHZ Hamiltonian derived above will be reconsidered for the open boundary conditions. For
the edge states, we deal with a semi-infinite plane, $y < 0$, so as only an edge solution of the form 
\begin{equation}
	\Psi _{\uparrow } \left (k_{x} ,y\right ) =\phi _{\lambda } \left (k_{x}\right ) e^{\lambda  y}\label{20x}
\end{equation}
is allowed (Re $\lambda  >0)$. The spatial dependence in the y-direction can be taken into account by applying
Peierls substitution: $k_{y} \rightarrow  -i \partial _{y}$ to $k_{y}$ in $\text{H}^{(0)} \left (\mathbf{k}\right )$. The solution $\Psi _{\downarrow }\left (k_{x} ,y\right )$ can easily be found by virtue of the time reversal operator $\Theta = -i \sigma_{y}\text{K}$  in  Eq.~(\ref{20x}) as $\Psi _{\downarrow }\left (k_{x} ,y\right )=\Theta \Psi _{\uparrow } \left (k_{x} ,y\right )$  where $\text{K}$ is  the complex conjugation operator.

Consequently, the secular equation gives two allowed values for $\lambda  :$ \begin{eqnarray*}
\lambda _{1 ,2}^{2} =k_{x}^{2} +F \pm \sqrt{F^{2} -\frac{\mathcal{M}^{2} -E^{2}}{\mathcal{B}_{ +} \mathcal{B}_{ -}}}  \nonumber
\end{eqnarray*}
with
\begin{eqnarray*}F =\frac{1}{2 \mathcal{B}_{ +} \mathcal{B}_{ -}} \left [\mathcal{A}^{2} -2 \left (\mathcal{M} \mathcal{B} +E \mathcal{D}\right )\right ]. \nonumber
\end{eqnarray*}
To find an edge state solution, the wave function must decay to zero when deviating from the boundary. Thus,
we adopt the Dirichlet boundary condition 
$\Psi _{\uparrow } \left (k_{x} ,y =0\right ) =\Psi _{\uparrow } \left (k_{x} ,y = -\infty \right ) =0$, then the general solution   in the presence of boundary is 
\begin{equation}
\Psi _{\uparrow } =\left (\begin{array}{c}\widetilde{c} (k_{x}) \\
\widetilde{d} (k_{x})\end{array}\right ) (e^{\lambda _{1} y} -e^{\lambda _{2} y})\label{23x}
\end{equation}
with $k_{x}$-dependent spinor coefficients $\widetilde{c} (k_{x})$ and $\widetilde{d} (k_{x})$. Since it is required that $\lambda$ should be positive to fullfill necessity of exponantially damping solution in Eq.~(\ref{23x}), one can follow the usual method to handle the energy depence of  
$\lambda _{1 ,2}$, and obtains
\begin{eqnarray*}
\lambda _{1 ,2} =\frac{1}{\sqrt{\mathcal{B}_{ +} \mathcal{B}_{ -}}} \left [\frac{\vert \mathcal{A}\vert }{2} \mp \sqrt{Z_{k_{x}}}\right ]\nonumber
\end{eqnarray*}
with
\begin{eqnarray*}
Z_{k_{x}} &  = & \left (\frac{\mathcal{A}^{2}}{4} -\frac{\mathcal{M}}{\mathcal{B}} \mathcal{B}_{ +} \mathcal{B}_{ -}\right )  -\frac{\mathcal{D} \vert \mathcal{A}\vert  \sqrt{\mathcal{B}_{ +} \mathcal{B}_{ -}}}{\mathcal{B}} k_{x} +\mathcal{B}_{ +} \mathcal{B}_{ -} k_{x}^{2}\nonumber
\end{eqnarray*}
which satisfies the conditions
\begin{eqnarray*}
\lambda _{1} \lambda _{2} =\frac{\mathcal{B} \mathcal{M} +\mathcal{D} E}{\mathcal{B}_{ +} \mathcal{B}_{ -}} -k_{x}^{2},\nonumber
\end{eqnarray*}
\begin{eqnarray*}
\lambda _{1} +\lambda _{2} =\frac{\mathcal{D} \mathcal{M} +\mathcal{B} E}{k_{x} \mathcal{B}_{ +} \mathcal{B}_{ -}}.\nonumber
\end{eqnarray*}

\begin{figure}[htp]
	\includegraphics[height=12cm,width=8.5cm]{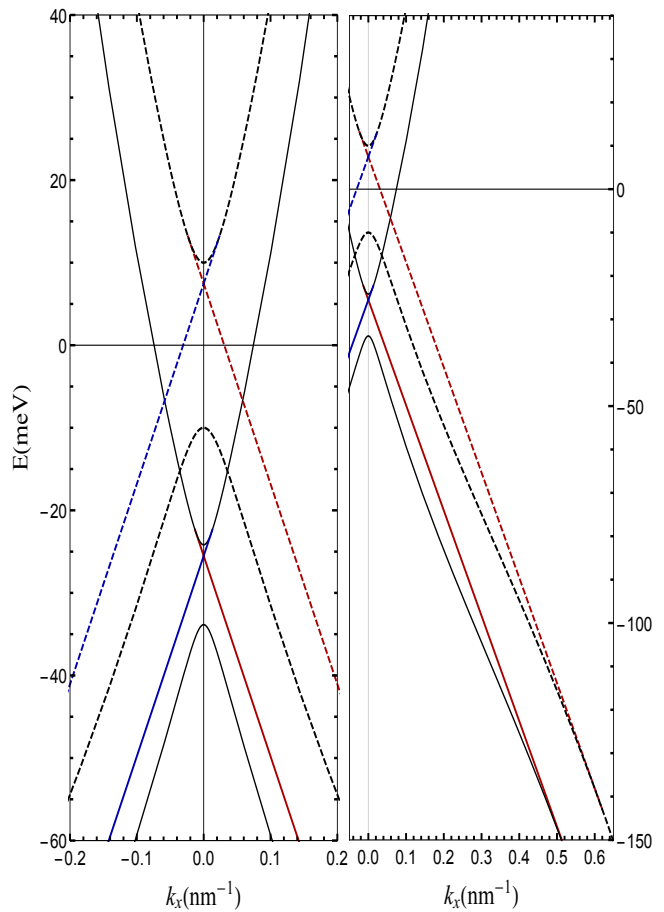}
	\caption{(Left Panel) Bulk and edge state dispersions obtained from our phonon-dressed BHZ model (solid lines) are compared with those obtained from the conventional one (dashed lines), i.e., in the absence of electron-SO phonon interaction. Here, we used $\mathrm{Si}\mathrm{C}$  substrate parameters given in Table~\ref{table1}, and $z= 0.3 \text{nm}$. (Right Panel) same as the left one, but to see where the HESs dive into the bulk, it is given  in large scales.}
	\label{fig1}
\end{figure}

\begin{figure}[htp]
	\includegraphics[height=12cm,width=8.5cm]{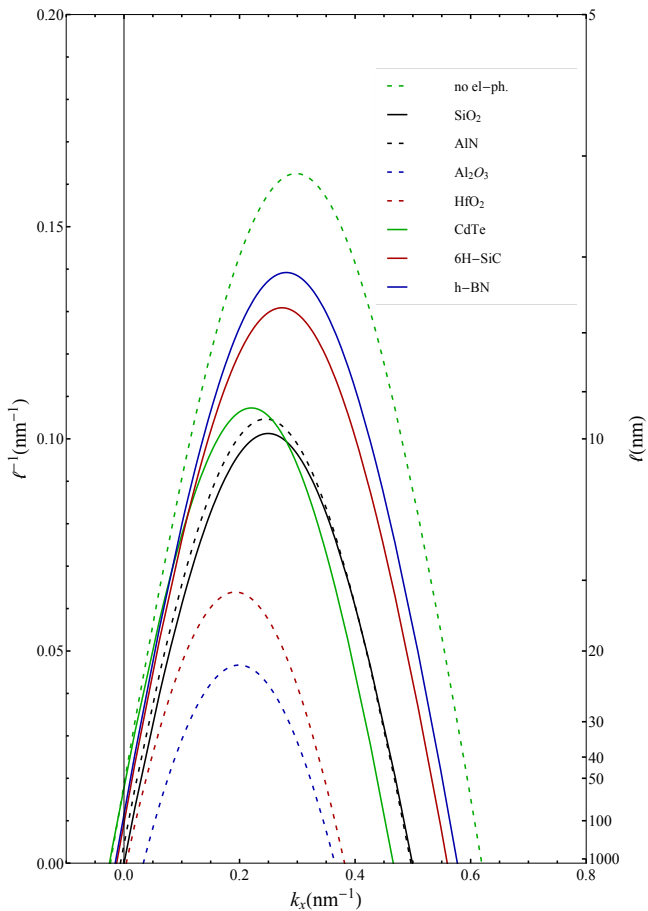}
	\caption{Inverse of the penetration depth length of the edge states for different substrates with $z =0.3\,\text{nm}$. Here, maxima of the curves correspond to the minima of the penetration depths. }\label{fig2}
\end{figure}

\begin{figure}[htp]
	\includegraphics[height=12cm,width=8.5cm]{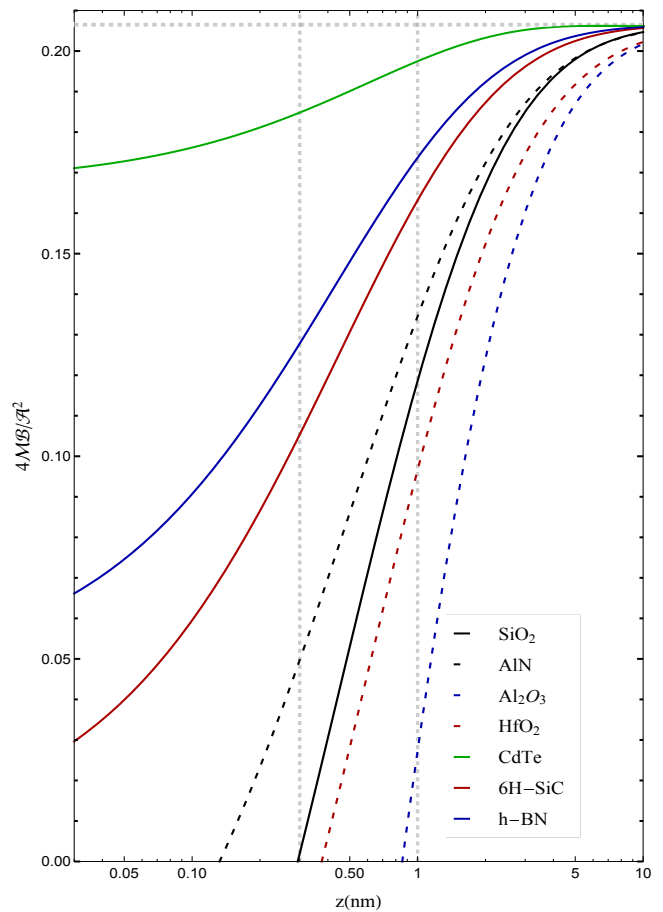}
	\caption{$4 \mathcal{M} \mathcal{B}/\mathcal{A}^{2}$ as a function of $z$ (in nanometers) for different substrates. }\label{fig3}
\end{figure}

\begin{figure}[htp]
	\includegraphics[height=12cm,width=8.5cm]{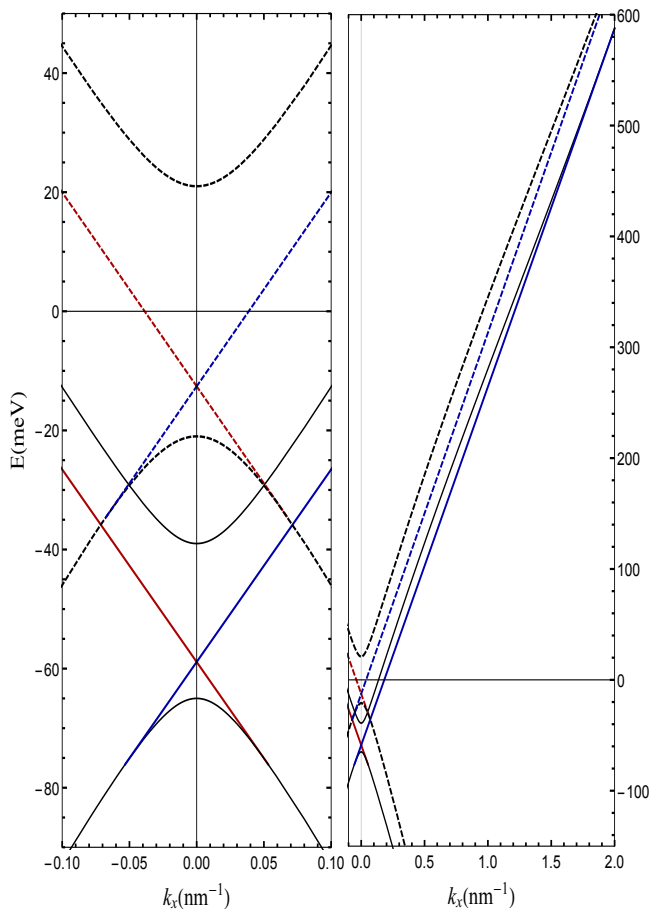}
	\caption{(Left panel) Bulk and edge state dispersion in the absence (dashed lines) and in the presence (solid lines) of electron-SO phonon interaction
		for a $\text{H}-\text{SiC}$ substrate for the parameters of $\text{BiSe}$, $\text{BiTe}$ films, $M = -0.021$$\text{eV}$, $D =7.5$$\text{eV} \mathring{A}$, $B = -12.5$$\text{eV} \mathring{A}^{2}$ and $\upsilon _{F} =6.16 \times 10^{5}$$\text{m/s}$.(Right Panel) same as the left one, but to see where the HESs dive into the bulk, it is given  in large scales.}\label{fig4}
\end{figure}

\begin{figure}[htb]
	\includegraphics[height=12cm,width=8.5cm]{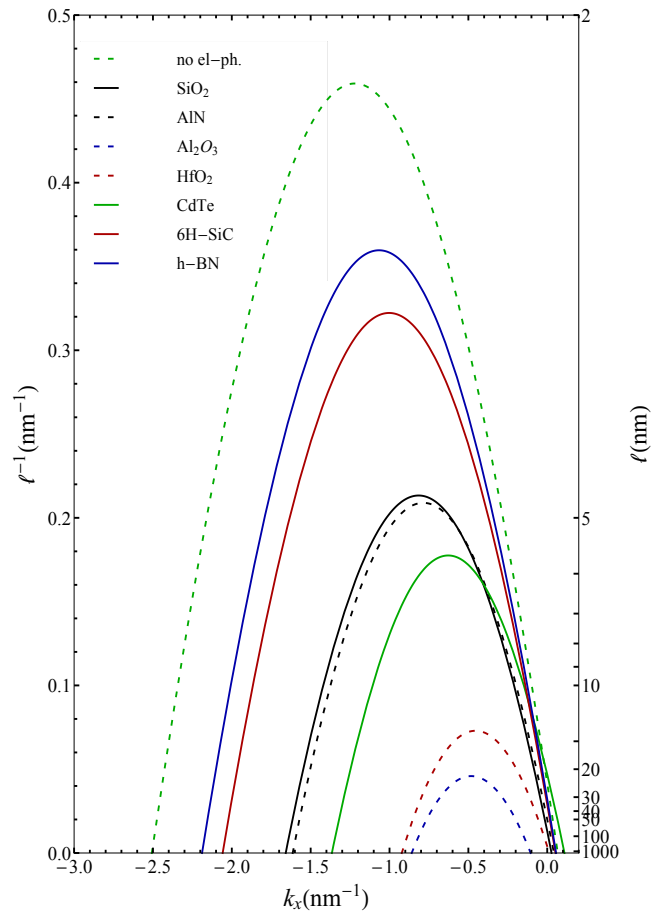}
	\caption{Penetration depth of the edge states for different substrates $\text{Bi}_{2}\text{Se}_{3}$ and $\text{Bi}_{3}\text{Te}_{3}$ thin films for $z =0 ,3$ nm, $M = -0.021$$\text{eV}$, $D =7.5$$\text{eV} \mathring{A}$, $B = -12.5$$\text{eV} \mathring{A}^{2}$ and $\upsilon _{F} =6.16 \times 10^{5}$$\text{m/s}$. }\label{fig5}
\end{figure}

\begin{table}[htb]
	\begin{center}\caption{Surface-optical phonon modes for different substrates $\text{SiO}_{2}$, $\text{AlN}$, $\text{Al}_{2}\text{O}_{3}$, $\text{HfO}_{2}$ (taken from Ref.~(\onlinecite{Fischetti2001})), $\text{CdTe}$ (taken from Ref.~(\onlinecite{Depaula1998})), $6\text{H}-\text{SiC}$(taken from Refs.~(\onlinecite{Fratini2008}-\onlinecite{Nienhaus1995})), and $\text{h-BN}$ (taken from Ref(\onlinecite{Perebeinos2010})).}
		\begin{tabular}{lllllllllc}\hline\hline
			$ $&$SiO_{2}$&$AlN$&$Al_{2}O_{3}$&$HfO_{2}$&$CdTe$&$6H-SiC$&$h-BN$\\ \hline
			$\epsilon_{0}$&3.9&9.14&12.53&22.0&10.23&9.7&5.1\\
			$\epsilon_{\omega}$&2.5&4.8&3.2&5.03&7.21&6.5&4.1\\
			$\omega_{SO,1}$&59.98&83.60&55.01&19.42&18.96&116&195\\
			$\omega_{SO,2}$&146.51&104.96&94.29&52.87&20.8&167.58&101\\
			$\beta$ &0.08&0.07&0.16&0.12&0.03&0.04&0.03\\
			\hline
			\hline\label{table1}
		\end{tabular}
	\end{center}
\end{table}

The energy spectrum of our phonon-dressed effective BHZ Hamiltonian  is given in FIG.~\ref{fig1} for a $\text{SiC}$ substrate. The bare material parameters we use here are from Ref.~\onlinecite{Konig2008},
$A =364 ,5$ $\text{meV nm}$, $B = -686$ $\text{meV nm}$, $M = -10$ $\text{meV }$, $D = -512$ $\text{meV nm}^{2}$, and  the surface optical phonon modes and the related dielectric constants of  substrates we used in this paper are summarized in Table.~\ref{table1}. In the left panel, bulk and edge state dispersions are given by using our phonon-dressed BHZ Hamiltonian, Eq.~(\ref{18})
and  Eq.~(\ref{19}), for the parameters of $\text{SiC}$.  In the right panel, 
all are given in wide scale to see where the HESs dive into the bulk. In the figure, while
the undressed bulk and edge states, i.e., states without electron-phonon interaction are given by dashed lines, dressed ones are represented by solid
lines. The edge states are displayed by using red (blue) curves for the spin-up (spin-down) case. Although we choose the energy offset $C$ to be equal to zero, it is easily seen from the figure that both valence and
conduction bands move down to deeper negative values , but  asymmetrically, just like an expected electronic behavior of graphene carriers in the presence of electron-phonon interaction\cite{Dubay2003,Pisana2007,KANDEMIRM2013,KANDEMIRM2014,KANDEMIRM2015,KANDEMIRA2017}. Moreover, 
in this proceses,  insulator-like behavior of the bulk and the metallic massless Dirac-like dispersion of the HESs are both preserved. However,  the slope of HESs is changed at the expense of decreasing gap term. It should be noticed that  the enhancement in the massive $D$ term due to the electron-SO phonon interaction that breaks the particle-hole symmetry gives rise to asymmetry between conduction and valence bands.
Therefore, the diving points of the HESs
to the bulk are modified depending on  the parameters of the substrate. As clearly seen from the right
panel of the FIG.\ref{fig1}, the region where the edge states exist is reduced in $k$ space compared with that found   in the absence of electron-SO phonon interaction. Hence, the penetration depth of the edge states becomes
longer in the presence of electron-SO phonon interaction. This means that penetration depth of the edge states into the bulk is not only the function of material
parameters but also the function of substrate parameters. Although the HESs are expected to be localized at the edge or at least near the edge, but in reality they are not, they
penetrate to the bulk. So their penetration depth length, $\ell $ which is expected to be of order of the lattice constant, and its control is recognized as an important issue in QSH systems
to be able to observe HESs\cite{Wada2011}. 

By assuming $\lambda _{1} >\lambda _{2}$, we plot behavior of inverse of the penetration depth length $\ell ^{ -1} =\lambda _{2}$ in FIG.~\ref{fig2} for different substrates. Its zeros, i.e., $k_{x}^{ \pm } =\mathcal{D} N \left [1 \pm \sqrt{1 +\left (\mathcal{B} \mathcal{M}\,/\mathcal{D}^{2} N^{2}\right )}\right ]/\mathcal{B}$, correspond to the points where HESs dive into the bulk in $k$ space with $N=\mathcal{A}/2  \sqrt{\mathcal{B}_{ +}\mathcal{B}_{ -}}$. In the absence of electron-SO phonon interaction the minimum of the penetration depth length occurs at $\left (k_{x}^{ +} +k_{x}^{ -}\right )/2 =0.30\, n m^{ -1}$ with $\ell _{\min } \sim 6.2\, \text{nm}$ which is compatible with that found in Ref~\onlinecite{Wada2011}.
The presence of electron-SO phonon interaction shifts the position of this minimum to a little bit smaller $k$ values,  due to the asymmetry  between conduction and valence bands caused by the massive character of parameter $\mathcal{D}$. Then $\ell _{\min }$ occurs over $10\, \text{nm}$, except that of $\text{SiC}$ and $\text{h-BN}$ substrates. This shows that the most suitable substrates are $\text{SiC}$ and $\text{h-BN}$ substrates with these material parameters to be able to observe HESs.

In the BHZ model, for real $\lambda $'s, HESs in the topological insulator regime exists only where $A^{2}/4 B^{2} \geq M/B \geq 0$.   In other words, $ M<0$  corresponds
to QSH regime, otherwise, i.e., $ M>0$ for a trivial state. To make a comparison of this region for a standard BHZ model and with that obtained by our
approach based on the phonon-dressed BHZ model,  we plot $4 \mathcal{M} \mathcal{B}/\mathcal{A}^{2}$ as a function of $z$ for different substrates in FIG.\ref{fig3}. This is just a number for a conventional $\text{HgTe}$
quantum well, i.e., $4 M B/A^{2} =0.207$, and shown by gray horizontal dashed line in FIG.~\ref{fig3}. Strikingly, this region is
getting smaller and smaller  for   substrates with high $ \beta$ values that indicate  high polarizability of the associated substrate, especially for experimentally accesible region of
$z$, i.e., around $3 -10\, \mathring{A}$. 

Substrate induced effects make  the quantity $4 M B/A^{2} $   $z$-dependent and critical $z >z_{c}$ occurs to fulfill the HESs criteria for substrates $\text{SiO}_{2}$, $\text{AlN}$, $\text{Al}_{2}\text{O}_{3}$ and $\text{HfO}_{2}$. For values of  $4 \mathcal{M} \mathcal{B}/\mathcal{A}^{2}$ close to zero,  it is impossible to observe HESs. On the contrary, substrates like $\text{SiC}$, $\text{h-BN}$ and $\text{CdTe}$ cover whole region without constraints on $z $ parameter.

These  phonon-dressed  material parameters can also be extended to derive an effective model for an ultrathin film of $\text{Bi}_{2}\text{Se}_{3}$ and $\text{Bi}_{3}\text{Te}_{3}$ compounds, e.g. films defined in Ref.~(\onlinecite{Lu2010}).
By taking into account the criteria $M/B >0$ for a gapless edge state, 
optimal numeric values for the model parameters can be found as $\upsilon _{F} =6.16 \times 10^{5}\; \mathrm{m}/\mathrm{s}$, $B = -12.5\; \mathrm{e} \mathrm{V} \mathring{A}^{2}$, $D =7.5\; \mathrm{e} \mathrm{V} \mathring{A}^{2}$, and $M = -0.021\; \mathrm{e} \mathrm{V}$ for a $L =32 \mathring{A}$ thicker quasi 2D topological insulator film from the Fig.~2 of Ref.~\onlinecite{Lu2010}.  It should be noted that the gap parameter value in this ultra
thin film geometry is almost two times larger than that of $\text{HgTe}/\text{CdTe}$ QWs. 
For this model, the bulk energy bands together with the associated  HESs are given in FIG.\ref{fig4} for $\text{SiC}$ substrate with $z=0.3,\text{nm}$. In this figure, we again  display the edge states by using red (blue) curves for the spin-up (spin-down) case. As in FIG. ~\ref{fig1},  although we choose the energy offset $C$ to be equal to zero,  both valence and
conduction bands move down to deeper negative values,  asymmetrically. Due to the large band gap, HESs dive to the bulk bands in large values of $k$ compared to those in FIG.~\ref{fig1} and thus survive in a wide range of $k$ in the Brillouin zone (BZ). This can be clearly seen from FIG. ~\ref{fig5} for different substrates.
In FIG.~\ref{fig5}
We plot the behavior of inverse of the penetration depth length $\ell ^{ -1} =\lambda _{2}$ in this figure for different substrates by using the material parameters of Ref.~(\onlinecite{Lu2010}). We notice that (i) the position of the minimum of the penetration depth length shifts to a little bit smaller $k$ values, due to the asymmetry  between conduction and valence bands caused by the massive character of parameter $\mathcal{D}$,  and (ii) it occurs over $5\, \text{nm}$, except that of $\text{SiC}$ and $\text{h-BN}$ substrates. In other words, HESs are well locaized around  $3\, \text{nm}$ in $\text{SiC}$ and $\text{h-BN}$ substrates with these material parameters.

\section{Conclusion}
In this, work, we show that the formation of HESs critically depends on the dielectric properties of substrates. Furthermore,
observation of these states on a given substrate depends  on the distance between the topological insulator and the substrate, as well as the
parameters of the substrate. 
Our results indicate that electron-SO phonon interactions have weak effects on the emergence of HESs in the case of $\text{h}-\text{BN}$ and  $6\text{H}-\text{SiC}$ due to their weak polarizability and high SO phonon frequencies. This can be understood  from the $\beta$ and  $\hbar  \omega$  dependence of the strength of the position dependent electron-phonon coupling paramater, i.e., $\alpha _{0} =e^{2} \beta /\sqrt{\hbar  \omega  \vert B_{ +}\vert }$. It is directly proportional to difference of the dielectric parameters of the material through $\beta$  and inverse square root of $\hbar  \omega $.  This quantity in the case of $\text{h}-\text{BN}$ and  $6\text{H}-\text{SiC}$  is less than that of other subtrates considered here.
So,  these substrates favor the emergence of HESs.
From our calculations, we also see that, for $\text{BiSe}$ and $\text{BiTe}$ thin films,  HESs survive in a wide range of $k$ in BZ for, in particular, $\text{h}-\text{BN}$ and  $6\text{H}-\text{SiC}$ substrates.  These compounds provide  more realistic model for observing HESs  in $\text{SiC}$ and $\text{h-BN}$ substrates which give rise to
well-locaized states around  $3\, \text{nm}$. 
Because of the fact that SO phonons induced by surface modes of the dielectric substrate may drastically  modify the electronic band topology of topological insulators together with the associated HESs, they can be used to tune the band gap and its sign of 2D topological insulators , and hence they can  play a critical role to drive the system from non-topological  state into a QSH phase.

\begin{acknowledgements}This work is supported by the Scientific and Technological Research Council of Turkey (T{\"U}B\.{I}TAK)
under the project number 115F421. 
\end{acknowledgements}

\appendix

\section{The matrix elements of $\text{H}^{(1)}$ and  $\text{H}^{(2)}$ Hamiltonians}
The diagonal matrix elements of the transformed Hamiltonian $\text{H}^{(1)}$ can be written as
\begin{eqnarray}
	\text{H}_{11}^{(1)} &  = & \sum _{\mathbf{q}}\left \{\left [\mathrm{M}_{\mathbf{q}}^{ \ast } (z) +\Omega ^{ -} f_{\mathbf{q}}\right ] b_{\mathbf{q}}^{ \dag } +\text{H.C.}\right \} \nonumber \\
\text{H}_{22}^{(1)}  &  = & \sum _{\mathbf{q}}\left \{\left [\mathrm{M}_{\mathbf{q}}^{ \ast } (z) +\Omega ^{ +} f_{\mathbf{q}}\right ] b_{\mathbf{q}}^{ \dag } +\text{H.C.}\right \} \label{A1}
\end{eqnarray}
together with non-diagonal ones
\begin{eqnarray}
\text{H}_{12}^{(1)}  &  = &  -A \sum _{\mathbf{q}}\mathbf{q}_{ +} \left (b_{\mathbf{q}}^{ \dag } f_{\mathbf{q}} +b_{\mathbf{q}} f_{\mathbf{q}}^{ \ast }\right ) \nonumber\\
\text{H}_{21}^{(1)} &  = &  -A \sum _{\mathbf{q}}\mathbf{q}_{ -} \left (b_{\mathbf{q}}^{ \dag } f_{\mathbf{q}} +b_{\mathbf{q}} f_{\mathbf{q}}^{ \ast }\right )
\end{eqnarray}
where $\Omega ^{ \pm } =\hbar  \omega  \pm \left (B \mp D\right ) \left [\mathbf{q}^{2} -2\, \mathbf{k} \cdot \mathbf{q}\,\left (1 -\eta \right )\right ]$\ \ \ and
finally the diagonal matrix elements of the transformed Hamiltonian of $\text{H}^{(2)}$ are
\begin{widetext}
\begin{eqnarray}
	\widetilde{H}_{1 1}^{2} &  = & \sum _{\mathbf{q}}\left [\Omega ^{ -} b_{\mathbf{q}}^{ \dag } b_{\mathbf{q}} -\left (B +D\right )\, f_{\mathbf{q}} \mathbf{q} \cdot \, \sum _{\mathbf{q}^{ \prime }}\mathbf{q}^{ \prime } f_{\mathbf{q}^{ \prime }}^{ \ast } b_{\mathbf{q}}^{ \dag } b_{\mathbf{q}^{ \prime }}\right ] \nonumber  \\
 &  &  -(B +D) \sum _{\mathbf{q}}\sum _{\mathbf{q}^{ \prime }}\mathbf{q} \cdot \mathbf{q}^{ \prime } \left \{b_{\mathbf{q}}^{ \dag } b_{\mathbf{q}^{ \prime }}^{ \dag } b_{\mathbf{q}} b_{\mathbf{q}^{ \prime }} +\left [f_{\mathbf{q}} f_{\mathbf{q}^{ \prime }} b_{\mathbf{q}}^{ \dag } f_{\mathbf{q}^{ \prime }}^{ \dag } +2 f_{\mathbf{q}^{ \prime }} b_{\mathbf{q}^{ \prime }}^{ \dag } b_{\mathbf{q}}^{ \dag } b_{\mathbf{q}} +\text{H.C.}\right ]\right \}\nonumber \\
\widetilde{H}_{2 2}^{2} &  = & \sum _{\mathbf{q}}\left [\Omega ^{ +} b_{\mathbf{q}}^{ \dag } b_{\mathbf{q}} +\left (B -D\right )\, f_{\mathbf{q}} \mathbf{q} \cdot \, \sum _{\mathbf{q}^{ \prime }}\mathbf{q}^{ \prime } f_{\mathbf{q}^{ \prime }}^{ \ast } b_{\mathbf{q}}^{ \dag } b_{\mathbf{q}^{ \prime }}\right ] \nonumber\\
 &  &  +(B -D) \sum _{\mathbf{q}}\sum _{\mathbf{q}^{ \prime }}\mathbf{q} \cdot \mathbf{q}^{ \prime } \left \{b_{\mathbf{q}}^{ \dag } b_{\mathbf{q}^{ \prime }}^{ \dag } b_{\mathbf{q}} b_{\mathbf{q}^{ \prime }} +\left [f_{\mathbf{q}} f_{\mathbf{q}^{ \prime }} b_{\mathbf{q}}^{ \dag } f_{\mathbf{q}^{ \prime }}^{ \dag } +2 f_{\mathbf{q}^{ \prime }} b_{\mathbf{q}^{ \prime }}^{ \dag } b_{\mathbf{q}}^{ \dag } b_{\mathbf{q}} +\text{H.C.}\right ]\right \} 
\end{eqnarray} 
\end{widetext}
together with non-diagonal ones
\begin{eqnarray}
\widetilde{H}_{1 2}^{2} &  = &  -A \sum _{\mathbf{q}}\mathbf{q}_{ +} b_{\mathbf{q}}^{ \dag } b_{\mathbf{q}} \nonumber\\
\widetilde{H}_{2 1}^{2} &  = &  -A \sum _{\mathbf{q}}\mathbf{q}_{ -} b_{\mathbf{q}}^{ \dag } b_{\mathbf{q}}.
\end{eqnarray}

\section[]{Substrate dependent parameters}
In this appendix, we give all phonon-dressed material parameters in Eq.~(\ref{19}) which are essential for Eq.~(\ref{18}) as
\begin{eqnarray}
C^{01} &  = & \frac{1}{2} \alpha _{0} \hbar  \omega  \left [\mathrm{Ci} \left (2 \overline{z}\right ) \sin  \left (2 \overline{z}\right ) +\frac{1}{2} \cos  (2 \overline{z}) \left (\pi  -2 \mathrm{Si} \left (2 \overline{z}\right )\right )\right ]\nonumber \\
\label{B1}\\
C^{02} &  = & C^{01} +\frac{1}{4 \sqrt{\pi }} \alpha _{0} \left (1 +\xi \right ) \hbar  \omega \, G_{1 ,3}^{3 ,1} \left (\overline{z}^{2}\left \vert \begin{array}{ccc}\, &  -\frac{1}{2} &  \\
0 & \frac{1}{2} & \frac{1}{2}\end{array}\right .\right ) \\
D^{01} &  = & \frac{1}{4 \sqrt{\pi }} \alpha _{0} \vert B_{ +}\vert  G_{1 ,3}^{3 ,1} \left (\overline{z}^{2}\left \vert \begin{array}{rrr}\, &  -\frac{1}{2} & \, \\
0 & \frac{1}{2} & \frac{3}{2}\end{array}\right .\right ) \\
D_{1}^{02} &  = & \frac{1}{12 \sqrt{\pi }} \alpha _{0} \vert B_{ +}\vert  G_{1 ,3}^{3 ,1} \left (\overline{z}^{2}\left \vert \begin{array}{rrr}\, &  -\frac{3}{2} & \, \\
0 & \frac{1}{2} & \frac{3}{2}\end{array}\right .\right ) \\
D_{2}^{02} &  = & \frac{1}{12 \sqrt{\pi }} \alpha _{0} \vert B_{ +}\vert  G_{1 ,3}^{3 ,1} \left (\overline{z}^{2}\left \vert \begin{array}{rrr}\, &  -\frac{1}{2} & \, \\
0 & \frac{1}{2} & \frac{5}{2}\end{array}\right .\right )\label{B5}
\end{eqnarray}
where $\mathrm{Ci}\left (x\right )$ is the cosine integral function 
$\mathrm{Ci} \left (x\right ) =\int _{ -z}^{\infty }d z \cos  (z)/z$, and similarly $\mathrm{Si} \left (z\right )$ is the sine integral function. Here, we have defined $D^{02}$ in terms of $D_{1}^{02}$and $D_{2}^{02}$ as $D^{02} =(2 +3 \xi ) D_{1}^{02} -D_{2}^{02}$, together with $\xi  =\vert B_{ -}\vert /\vert B_{ +}\vert $.

\end{document}